\documentclass[12pt]{iopart}
\usepackage{amsfonts}
\usepackage[english]{babel}
\usepackage{euscript}
\usepackage{color}

\begin{document}

\title[Recursion Operator via Cartan's Equivalence Method]{%
A Recursion Operator for the Universal Hierarchy Equation via
Cartan's Method of Equivalence
}

\author{Oleg I. Morozov}

\address{
Institute of Mathematics and Statistics, University of Troms\o, Troms\o 
\, 90-37, Norway
\\ 
E-mail:  Oleg.Morozov{\symbol{64}}uit.no}

\begin{abstract}
We apply Cartan's method of equivalence to find a B\"ack\-lund auto\-trans\-for\-ma\-ti\-on 
for the tangent covering of the universal hierarchy equation. The trans\-for\-ma\-ti\-on provides a recursion 
operator for symmetries of this equation.
\end{abstract}



\maketitle

\section{Introduction}

Recursion operators  provide an important tool for the study of  nonlinear partial dif\-fe\-ren\-ti\-al equations {\sc pde}s. 
They are linear operators acting  on a symmetries of a {\sc pde} and generating in\-fi\-ni\-te hierarchies of new 
(nonlocal) symmetries. The presence of infinite series of symmetries allows one to apply different techniques to 
study a given {\sc pde}, see \cite{Vinogradov1984,Ibragimov1985,KrasilshchikLychaginVinogradov1986,Olver1993,VK1999} 
and references therein. Typically, recursion operators are con\-si\-de\-red as integro-differential operators,
\cite{Olver1977,
Fuchssteiner1979,
ZakharovKonopelchenko1984,
FokasSantini1986,
Fokas1987,
FokasSantini1988,
SantiniFokas1988,
BlumanKumei1989,
Olver1993,
GursesKarasuSokolov1999,
Sergeyev2000,
SandersWang2001,
Wang2002},
al\-though this interpretation  is accompanied by a number of difficulties, e.g., discussed in
\cite{Guthrie1994,KrasilshchikVinogradov1984}.
Another definition is proposed in 
\cite{Papachristou1990,Guthrie1994,KrasilshchikKersten1994,KrasilshchikKersten1995}
(see also \cite{PapachristouHarrison2010} and references therein) 
and developed in
\cite{
Marvan1996,
Sergeyev2000,
MarvanSergyeyev2003,
Marvan2004a,
Marvan2004b,
MarvanPoborzhil2008,
Marvan2010}. It treats recursion operators as the B\"acklund auto\-trans\-for\-ma\-ti\-ons of
the tangent (or linearized) coverings of the {\sc pde}s. In \cite{MarvanSergyeyev2012}, M. Marvan and A. Sergyeyev 
proposed  the method for constructing recursion operators of {\sc pde}s of any di\-men\-si\-on from their linear 
coverings of a special form. By this method they found re\-cur\-si\-on operators for a number of {\sc pde}s of 
physical and geometrical significance.

In the present paper we adapt the technique of \cite{Morozov2009b} for the problem of finding recursion operators
of {\sc pde}s. This technique is based on \'Elie Cartan's structure theory of Lie pseudo-groups, 
\cite{Cartan1,Cartan4,Vasilieva1972,Stormark2000},  and allows one to find coverings and B\"acklund transformations for nonlinear 
{\sc pde}s by means of contact integrable extensions of their symmetry pseudo-groups. 
We consider  the universal hierarchy equation \cite{Pavlov2003,MartinezAlonsoShabat2002,MartinezAlonsoShabat2004}
\begin{equation}
u_{yy} = u_y\, u_{tx}-u_x\,u_{ty}.
\label{universal_hierarchy_eq}
\end{equation}
In accordance with \cite[\S 5]{KrasilshchikVerbovetskyVitolo2011} we identify the tangent covering 
for (\ref{universal_hierarchy_eq}) with the system of {\sc pde}s constituted by (\ref{universal_hierarchy_eq}) and 
the defining equation for its symmetries
\begin{equation}
v_{yy} = u_y\, v_{tx} + v_y\, u_{tx}-u_x\,v_{ty} - v_x\,u_{ty}.
\label{linearized_universal_hierarchy_eq}
\end{equation}
The standard procedures of \'Elie Cartan's method of equivalence give Maurer--Cartan forms and the structure 
equations of the pseudo-group of contact symmetries for  system (\ref{universal_hierarchy_eq}), (\ref{linearized_universal_hierarchy_eq}). Then we find 
a contact integrable extension for these structure equations. The corresponding contact form provides the 
B\"acklund auto-transformation for Eq. (\ref{linearized_universal_hierarchy_eq}). This transformation defines a 
recursion operator for symmetries of Eq. (\ref{universal_hierarchy_eq}) and its in\-ver\-se operator.

\section{Preliminaries}\label{Preliminaries_section}

\subsection{Coverings of {\sc pde}s}

Let $\pi \colon \mathbb{R}^n \times \mathbb{R}^m \rightarrow \mathbb{R}^n$,
$\pi \colon (x^1, \dots, x^n, u^1, \dots, u^m) \mapsto (x^1, \dots, x^n)$, be a trivial bundle, and $J^\infty(\pi)$
be the bundle of its jets of the infinite order. The local coordinates on $J^\infty(\pi)$ are $(x^i,u^\alpha,u^\alpha_I)$,
where $I=(i_1, \dots, i_n)$ is a multi-index, and for every local section
$f \colon \mathbb{R}^n \rightarrow \mathbb{R}^n \times \mathbb{R}^m$ of $\pi$ the corresponding infinite jet $j_\infty(f)$ is
a section $j_\infty(f) \colon \mathbb{R}^n \rightarrow J^\infty(\pi)$ such that
$u^\alpha_I(j_\infty(f))
=\displaystyle{\frac{\partial ^{\#I} f^\alpha}{\partial x^I}}
=\displaystyle{\frac{\partial ^{i_1+\dots+i_n} f^\alpha}{(\partial x^1)^{i_1}\dots (\partial x^n)^{i_n}}}$.
We put $u^\alpha = u^\alpha_{(0,\dots,0)}$. Also, in the case of $n=3$, $m=1$ we denote $x^1 = t$, $x^2= x$, 
$x^3= y$, and $u^1_{(i,j,k)}=u_{{t \dots t}{x \dots x}{y \dots y}}$  
with $i$  times $t$, $j$  times $x$, and $k$  times $y$. 

The vector fields
\[
D_{x^k} = \frac{\partial}{\partial x^k} + \sum \limits_{\# I \ge 0} \sum \limits_{\alpha = 1}^m
u^\alpha_{I+1_{k}}\,\frac{\partial}{\partial u^\alpha_I},
\qquad k \in \{1,\dots,n\},
\]
$(i_1,\dots, i_k,\dots, i_n)+1_k = (i_1,\dots, i_k+1,\dots, i_n)$,  are called {\it total derivatives}.  They commute everywhere on
$J^\infty(\pi)$:  $[D_{x^i}, D_{x^j}] = 0$.

\vskip 10 pt

The {\it evolutionary differentiation} associated to an arbitrary vector-valued smooth function
$\varphi \colon J^\infty(\pi) \rightarrow \mathbb{R}^m $ is the vector
field
\begin{equation}
\mathbf{E}_{\varphi} = \sum \limits_{\# I \ge 0} \sum \limits_{\alpha = 1}^m D_I(\varphi^\alpha)\,\frac{\partial}{\partial u^\alpha_I},
\label{evolution_differentiation}
\end{equation}
with $D_I=D_{(i_1,\dots\,i_n)} =D^{i_1}_{x^1} \circ \dots \circ D^{i_n}_{x^n}$.

\vskip 10 pt

A system of {\sc pde}s $F_r(x^i,u^\alpha_I) = 0$, $\# I \le s$, $r \in \{1,\dots, R\}$, of the order 
$s \ge 1$ with $R \ge 1$ defines the submanifold
$\EuScript{E} = \{(x^i,u^\alpha_I) \in J^\infty(\pi) \,\,\vert\,\, D_K(F_r(x^i,u^\alpha_I)) = 0, \,\, \# K \ge 0\}$
in $J^\infty(\pi)$.

\vskip 10 pt

A function $\varphi \colon J^\infty(\pi) \rightarrow \mathbb{R}^m$ is called a {\it (generator of an infinitesimal) symmetry} of
$\EuScript{E}$ when $\mathbf{E}_{\varphi}(F) = 0$ on $\EuScript{E}$. The symmetry $\varphi$ is a solution to the {\it defining system}
\begin{equation}
\ell_{\EuScript{E}}(\varphi) = 0,
\label{defining_eqns}
\end{equation}
where $\ell_{\EuScript{E}} = \ell_F \vert_{\EuScript{E}}$ with the matrix differential operator
\[
\ell_F = \left(\sum \limits_{\# I \ge 0}\frac{\partial F_r}{\partial u^\alpha_I}\,D_I\right).
\]

\vskip 10 pt

Denote $\mathcal{W} = \mathbb{R}^\infty$ with  coordinates $w^s$, $s \in  \mathbb{N} \cup \{0\}$. Locally,
an (infinite-di\-men\-si\-o\-nal)  {\it differential covering} of $\EuScript{E}$ is a trivial bundle $\tau \colon J^\infty(\pi) \times \mathcal{W} \rightarrow J^\infty(\pi)$
equipped with the {\it extended total derivatives}
\begin{equation}
\widetilde{D}_{x^k} = D_{x^k} + \sum \limits_{ s =0}^\infty
T^s_k(x^i,u^\alpha_I,w^j)\,\frac{\partial }{\partial w^s}
\label{extended_derivatives}
\end{equation}
such that $[\widetilde{D}_{x^i}, \widetilde{D}_{x^j}]=0$ for all $i \not = j$ whenever $(x^i,u^\alpha_I) \in \EuScript{E}$. We define
the partial derivatives of $w^s$ by  $w^s_{x^k} =  \widetilde{D}_{x^k}(w^s)$.  This yields the system of
{\it covering equations}
\begin{equation}
w^s_{x^k} = T^s_k(x^i,u^\alpha_I,w^j).
\label{WE_prolongation_eqns}
\end{equation}
This over-determined system of {\sc pde}s is compatible whenever $(x^i,u^\alpha_I) \in \EuScript{E}$.

\vskip 10 pt

Denote by $\widetilde{\mathbf{E}}_\varphi$  the result of substitution  $\widetilde{D}_{x^k}$ for
$D_{x^k}$ in (\ref{evolution_differentiation}). A  {\it shadow of  nonlocal symmetry} of $\EuScript{E}$
corresponding to the covering $\tau$ with the extended total derivatives (\ref{extended_derivatives}),
or $\tau$-{\it shadow}, is a function $\varphi \in C^\infty(\EuScript{E} \times \mathcal{W})$ such that
\begin{equation}
\widetilde{\mathbf{E}}_\varphi(F) = 0
\label{shadow_eqn}
\end{equation}
is a consequence of equations $D_K(F)=0$ and (\ref{WE_prolongation_eqns}).
A {\it nonlocal symmetry} of $\EuScript{E}$ cor\-res\-pon\-ding to the covering $\tau$ (or $\tau$-{\it symmetry})
is the vector field
\begin{equation}
\widetilde{\mathbf{E}}_{\varphi,A} = \widetilde{\mathbf{E}}_{\varphi}
+\sum \limits_{s=0}^\infty A^s\,\frac{\partial}{\partial w_s},
\label{extended_nonlocal_evolution_differentiation}
\end{equation}
with $A^s \in C^\infty(\EuScript{E} \times \mathcal{W})$ such that
$\varphi$ satisfies to  (\ref{shadow_eqn})  and
\begin{equation}
\widetilde{D}_k(A^s)=\widetilde{\mathbf{E}}_{\varphi,A}(T^s_k)
\label{obstruction_eqn}
\end{equation}
for $T^s_k$ from (\ref{extended_derivatives}), see \cite[Ch. 6, \S 3.2]{VK1999}.

\vskip 10 pt
\noindent
{\sc remark} 1.
In general, not every $\tau$-shadow  corresponds to a $\tau$-symmetry, since Eqns. (\ref{obstruction_eqn})
provide an obstruction for existence of (\ref{extended_nonlocal_evolution_differentiation}). But for any $\tau$-shadow
$\varphi$ there exists a covering $\tau_{\varphi}$ and a nonlocal $\tau_{\varphi}$-symmetry whose
$\tau_{\varphi}$-shadow coincides with $\varphi$, see \cite[Ch. 6, \S 5.8]{VK1999}.

\vskip 10 pt

A {\it recursion operator} $\mathcal{R}$   for $\EuScript{E}$ is a $\mathbb{R}$-linear map such that for each 
(local or nonlocal)  symmetry $\varphi$ of  $\EuScript{E}$ the function $\mathcal{R}(\varphi)$ is a (local or 
nonlocal) symmetry of $\varphi$ of  $\EuScript{E}$.

\vskip 10 pt

The tangent covering for {\sc pde} $\EuScript{E}$ is defined as follows, \cite{KrasilshchikVerbovetskyVitolo2011}.
Consider the trivial bundle
$\sigma \colon J^\infty(\pi)\times \mathcal{V} \rightarrow J^\infty(\pi)$
with  coordinates $v^\alpha_I$, $\# I \ge 0$, on the fiber $\mathcal{V}$ equipped with the extended total derivatives
\[
\hat{D}_{x^k} = D_{x^k}+\sum \limits_{\# I \ge 0} \sum \limits_{\alpha = 1}^m
v^\alpha_{I+1_k}\,\frac{\partial}{\partial v^\alpha_I}.
\]
Then for $\hat{D}_I= \hat{D}^{i_1}_{x^1} \circ \dots \circ \hat{D}^{i_n}_{x^n}$ define
\[
\hat{\ell}_F = \left(
\sum \limits_{\# I \ge 0}\frac{\partial F_r}{\partial u^\alpha_I}\,\hat{D}_I
\right).
\]
and put
\[
\fl
\EuScript{T(E)} =
\{
(x^i,u^\alpha_i,v^\alpha_I) \in J^\infty(\pi)\times \mathcal{V}
\,\,\,\vert \,\,\,
D_K(F(x^i,u^\alpha_I))=0,\,\,
\hat{D}_K (\hat{\ell}_F(v^\alpha))=0,\,\,
\# K \ge 0
\}.
\]
The {\it  tangent covering} is the restriction of $\sigma$ to $\EuScript{T(E)}$. A section
$\varphi \colon \EuScript{E} \rightarrow \EuScript{T(E)}$ of the tangent covering is a symmetry of $\EuScript{E}$.
The extended total derivatives of this covering are
$\widetilde{D}_{x^k} = \hat{D}_{x^k}\vert_{\EuScript{T(E)}}$.

\vskip 10 pt
\noindent
{\sc example} 1.\,\,\,
We write Eq. (\ref{universal_hierarchy_eq}) in the form 
$u_{yy} - u_y\,u_{tx}+u_x\,u_{ty} = 0$. Then we have
\[
\ell_F(\varphi) = 
D_y^2(\varphi)
-u_yD_tD_x(\varphi)
-u_{tx}D_y(\varphi)
+u_xD_tD_y(\varphi)
+u_{ty}D_x(\varphi)
\]
and
\[
\hat{\ell}_F(v) = v_{(0,0,2)}-u_y v_{(1,1,0)}-u_{tx} v_{(0,0,1)}+u_x v_{(1,0,1)}+u_{ty} v_{(0,1,0)}.
\]
The fiber of the tangent covering has local coordinates
$v_{(i,j,0)}$ and $v_{(i,j,1)}$. The extended total derivatives of the tangent covering are
\[
\fl
\left\{
\begin{array}{lcl}
\widetilde{D}_t &=&
\displaystyle{
D_t
+ \sum \limits_{i=0}^\infty \sum \limits_{j=0}^\infty
\left(
v_{(i+1,j,0)}\,\frac{\partial}{\partial v_{(i,j,0)}}
+ v_{(i+1,j,1)}\,\frac{\partial}{\partial v_{(i,j,1)}}
\right),
}
\\
\widetilde{D}_x &=&
\displaystyle{
D_x
+ \sum \limits_{i=0}^\infty \sum \limits_{j=0}^\infty
\left(
v_{(i,j+1,0)}\,\frac{\partial}{\partial v_{(i,j,0)}}
+ v_{(i,j+1,1)}\,\frac{\partial}{\partial v_{(i,j,1)}}
\right),
}
\\
\widetilde{D}_y &=&
\displaystyle{
D_y
+ \sum \limits_{i=1}^\infty \sum \limits_{j=1}^\infty
v_{(i,j,1)}\,\frac{\partial}{\partial v_{(i,j,0)}}
}
\\
&&
\displaystyle{
\phantom{D_y}
+ \sum \limits_{i=1}^\infty \sum \limits_{j=1}^\infty
\widetilde{D}_t^i\widetilde{D}_x^j(u_y v_{(1,1,0)}+u_{tx} v_{(0,0,1)}-u_x v_{(1,0,1)}-u_{ty} v_{(0,1,0)})\,\frac{\partial}{\partial v_{(i,j,1)}}
}.
\end{array}
\right.
\]

\vskip 10 pt
\noindent
{\sc remark}  2.    Abusing the notation, we write  $v_{{t \dots t}{x \dots x}{y \dots y}}$  
with $i$  times $t$, $j$  times $x$, $k$  times $y$ instead of $v_{(i,j,k)}$ in what follows.
Also, we identify the tangent covering with the coverings equations 
$\hat{\ell}_F(v)=0$, i.e., with Eqs. (\ref{universal_hierarchy_eq}), (\ref{linearized_universal_hierarchy_eq})

\subsection{Cartan's structure theory of Lie pseudo-groups}

Let $M$ be a manifold of dimension $n$. A {\it local diffeomorphism} on $M$ is a diffeomorphism 
$\Phi \colon \EuScript{U} \rightarrow \hat{\EuScript{U}}$ of two open subsets of $M$. A {\it pseudo-group} 
$\mathfrak{G}$ on $M$ is a collection of local dif\-feo\-mor\-phisms of $M$, which is closed under composition 
{\it when defined}, contains an identity and is closed under inverse. A {\it Lie pseudo-group} is a pseudo-group 
whose diffeomorphisms are local analytic solutions of an involutive system of partial differential equations called 
{\it defining system}.

\'Elie Cartan's approach to Lie pseudo-groups is based on a possibility to characterize transformations from 
a pseudo-group in terms of a set of invariant differential 1-forms called {\it Maurer--Cartan} ({\sc mc}) {\it forms}.
In a general case, {\sc mc} forms $\omega^1$, ... , $\omega^m$ of an infinite-dimensional Lie pseudo-group 
$\mathfrak{G}$ are defined on a direct product $M \times \tilde{M} \times G$, where $\tilde{M}$ is the coordinate 
space of parameters of prolongation, \cite[Ch. 12]{Olver1995}, $G$ is a finite-dimensional Lie group, and  
$m = \dim \,M +\dim \,\tilde{M}$. The forms $\omega^i$ are independent and include differentials of coordinates on 
$M \times \tilde{M}$ only, while their coefficients depend also on coordinates of $G$. These  forms characterize the 
pseudo-group $\mathfrak{G}$ in the following sense: a local diffeomorphism 
$\Phi \colon \EuScript{U} \rightarrow \hat{\EuScript{U}}$ on $M$ belongs to $\mathfrak{G}$ whenever there exists a 
local diffeomorphism $\Psi \colon \EuScript{W} \rightarrow \hat{\EuScript{W}}$ on $M \times \tilde{M}\times G$ such 
that $\rho \circ \Psi = \Phi \circ \rho$ for the projection 
$\rho \colon M \times \tilde{M} \times G \rightarrow M$ and the forms $\omega^j$ are invariant w.r.t. $\Psi$, that 
is,
\begin{equation}
\Psi^{*} \left(\omega^i\vert {}_{\hat{\EuScript{W}}} \right) 
= \omega^i\vert {}_{\EuScript{W}}.
\label{Phi_omega}
\end{equation}
Expressions for the exterior differentials of the forms $\omega^i$ in terms of themselves give Cartan's 
{\it structure equations} of $\mathfrak{G}$:
\begin{equation}
d \omega^i = A_{\gamma j}^i \,\pi^\gamma \wedge \omega^j + B_{jk}^i\,\omega^j \wedge \omega^k,
\qquad B_{jk}^i = - B_{kj}^i.
\label{SE} 
\end{equation}
The forms $\pi^\gamma$, $\gamma \in \{1,...,\dim \, G\}$, are linear combinations of {\sc mc} forms of the Lie group 
$G$ and the forms $\omega^i$. The coefficients $A_{\gamma j}^i$ and $B_{jk}^i$ are either constants or functions of 
a set of invariants $U^\kappa \colon M \rightarrow \mathbb{R}$, $\kappa \in \{1,...,l\}$, 
$l < \dim\, M$, of the pseudo-group $\mathfrak{G}$, so 
$\Phi^{*} \left(U^{\kappa}\vert {}_{\hat{\EuScript{U}}} \right) = U^{\kappa}\vert {}_{\EuScript{U}}$
for every $\Phi \in \mathfrak{G}$. In the latter case, the differentials of $U^\kappa$ are invariant 1-forms, so 
they are linear combinations of the forms $\omega^j$,
\begin{equation}
d U^\kappa = C_j^\kappa\,\omega^j,
\label{dUs}
\end{equation} 
where the coefficients $C_j^\kappa$ depend on the invariants $U^1$, ..., $U^l$ only.

Eqs. (\ref{SE}) must be compatible in the following sense: we have  
\begin{equation}
d(d \omega^i) = 0 = d \left(A_{\gamma j}^i \,\pi^\gamma \wedge \omega^j + B_{jk}^i\,\omega^j \wedge \omega^k \right),
\label{compatibility_conditions_SE} 
\end{equation}
therefore there must exist expressions 
\begin{equation}
d \pi^\gamma = W_{\lambda j}^\gamma\, \chi^\lambda \wedge \omega^j 
+ X_{\beta \epsilon}^\gamma\,\pi^\beta\wedge\pi^\epsilon
+Y_{\beta j}^\gamma\,\pi^\beta \wedge \omega^j
+Z_{jk}^\gamma\,\omega^j \wedge \omega^k
\label{prolonged_SE}
\end{equation}
with some additional 1-forms $\chi^\lambda$  such that the right-hand side of (\ref{compatibility_conditions_SE}) 
is identically equal to zero after substituting for (\ref{SE}), (\ref{dUs}), and (\ref{prolonged_SE}). 
Also, from (\ref{dUs}) it follows that
the right-hand side of the equation
\begin{equation}
d(d U^\kappa) = 0 = d(C_j^\kappa\,\omega^j)
\label{compatibility_conditions_dUs}
\end{equation}
must be identically equal to zero after substituting for (\ref{SE}) and (\ref{dUs}).

The forms $\pi^\gamma$ are not invariant w.r.t. the pseudo-group $\mathfrak{G}$. Respectively, the structure equations (\ref{SE}) are not changing when replacing 
$\pi^\gamma \mapsto \pi^\gamma + z^\gamma_j\,\omega^j$ for certain parametric coefficients $z^\gamma_j$. The dimension $r^{(1)}$ of the linear space of these coefficients satisfies the fol\-lo\-wing inequality
\begin{equation}
r^{(1)} \le n\,\dim \, G  -  \sum \limits_{k=1}^{n-1} (n-k)\,s_k, 
\label{CT}
\end{equation}
where the {\it reduced characters} $s_k$ are defined by the formulas
\begin{eqnarray}
s_1 &=& \max \limits_{u_1\in \mathbb{R}^n}\, \mathrm{rank}\,\, \mathbb{A}_1(u_1),
\nonumber
\\
s_k &=& \max \limits_{u_1,...,u_k \in \mathbb{R}^n}\, \mathrm{rank}\,\, \mathbb{A}_k(u_1,...,u_k) - 
\sum \limits_{j=1}^{k-1} s_j, \qquad k \in \{1, ... , n-1\},
\nonumber
\\
s_n &=& \mathrm{dim}\, G - \sum \limits_{j=1}^{n-1} s_j,
\nonumber
\end{eqnarray}
with the  matrices $\mathbb{A}_k$ inductively defined by
\[
\fl
\mathbb{A}_1(u_1) = \left(A^i_{\gamma j} \,u^j_1\right),
\qquad \mathbb{A}_l(u_1,...,u_l) = \left(
\begin{array}{c}
\mathbb{A}_{l-1}(u_1,...,u_{l-1})
\\
A^i_{\gamma j} \,u^j_l
\end{array}
\right),
\qquad l \in \{2, ... n-1\},
\]
see \cite[\S 5]{Cartan1}, \cite[Def. 11.4]{Olver1995} for the full discussion. The system of forms $\omega^k$ is {\it involutive}  when both sides of (\ref{CT}) are equal, \cite[\S 6]{Cartan1}, \cite[Def. 11.7]{Olver1995}.

Cartan's fundamental theorems, \cite[\S\S 16, 22--24]{Cartan1}, \cite{Cartan4}, 
\cite[\S\S 16, 19, 20, 25,26]{Vasilieva1972}, \cite[\S\S 14.1--14.3]{Stormark2000}, state that for a Lie 
pseudo-group there exists a set of {\sc mc} forms whose structure equations satisfy the compatibility and 
involutivity conditions; conversely, if Eqs. (\ref{SE}), (\ref{dUs}) meet the compatibility conditions 
(\ref{compatibility_conditions_SE}), (\ref{compatibility_conditions_dUs}) and the involutivity con\-di\-tion, then 
there exists a collection of 1-forms $\omega^1$, ... , $\omega^m$ and functions $U^1$, ... , $U^l$ which satisfy 
(\ref{SE}) and (\ref{dUs}).  Eqs. (\ref{Phi_omega}) then define local diffeomorphisms from a Lie pseudo-group.

\section{Symmetry pseudo-group of the  tangent covering of the universal hierarchy equation} 

Using the procedures of \'Elie Cartan's method of equivalence, 
\cite{Cartan1,Cartan2,Cartan3,Cartan4,Vasilieva1972,Gardner,Kamran,Olver1995,FelsOlver,Morozov2002,Morozov2006}, we 
find the Maurer--Cartan forms and their structure equations for the symmetry pseudo-group of system 
(\ref{universal_hierarchy_eq}), (\ref{linearized_universal_hierarchy_eq}). The full set of involutive structure 
equations for this pseudo-group consists of two parts:
\begin{eqnarray}
\fl
d\theta_0 &=&
\theta_0 \wedge (\theta_3-\xi^1-U_2 \, \xi^2-U_1 \, \xi^3-\theta_{12})
+\xi^1 \wedge \theta_1
+\xi^2 \wedge \theta_2
+\xi^3 \wedge \theta_3,
\nonumber
\\
\fl
d\theta_1 &=&
\eta_1 \wedge \theta_1
+\theta_0 \wedge (\theta_{13}+(U_1+1) \, \xi^2+\xi^3)
+\xi^1 \wedge \theta_{11}
+\xi^2 \wedge \theta_{12}
+\xi^3 \wedge \theta_{13},
\nonumber
\\
\fl 
d\theta_2 &=&
\theta_2 \wedge \eta_1
+\xi^1 \wedge (\theta_{12} - \theta_{13})  
+\xi^2 \wedge \theta_{22}
+\xi^3 \wedge (\theta_{23} - \theta_2),
\nonumber
\\
\fl
d\theta_3 &=&
\xi^1 \wedge \theta_{13}
+\xi^2 \wedge \theta_{23}
+\xi^3 \wedge \theta_{12},
\nonumber
\\
\fl
d\xi^1 &=&
(\theta_{12} - \eta_1 - \theta_3 + U_2 \, \xi^2 + U_1\,\xi^3) \wedge \xi^1,
\nonumber
\\
\fl
d\xi^2 &=&
(\eta_1+\theta_{12} + \xi^1 + (U_1+1)  \, \xi^3)  \wedge \xi^2,
\nonumber
\\
\fl
d\xi^3 &=&
(\theta_2 +\xi^1)\wedge \xi^2
+(\theta_{12}+\xi^1 + U_2\, \xi^2) \wedge \xi^3,
\nonumber
\\
\fl
d\theta_{11} &=&
\eta_1 \wedge (\xi^2+2 \, \theta_{11})
+\eta_2 \wedge (\theta_0+\xi^3)
+\eta_5 \wedge \xi^1
+\theta_0 \wedge (\theta_{13}+2 \, (U_1+1) \, \xi^2+2 \, \xi^3)
\nonumber
\\
\fl
&&
+\theta_1 \wedge (\theta_{13}+(U_1+1) \, \xi^2+\xi^3)
+\theta_3 \wedge (\xi^2+\theta_{11})
+\xi^2 \wedge (2 \, \theta_{12}+\theta_{13}-U_2 \, \theta_{11})
\nonumber
\\
\fl
&&
+\xi^3 \wedge (\theta_{13} -U_1 \, \theta_{11})
+\theta_{11} \wedge \theta_{12},
\nonumber
\\
\fl
d\theta_{12} &=&
\eta_1 \wedge \xi^1
+(\eta_3 +\theta_2-U_2 \, \theta_{13}) \wedge \xi^3
+\eta_4 \wedge \xi^2
+(\theta_3-\theta_{12}) \wedge (\xi^1-(U_1+1) \, \xi^3),
\nonumber
\\
\fl
d\theta_{13} &=&
(\eta_1+\theta_3) \wedge (\xi^3+\theta_{13})
+\eta_2 \wedge \xi^1
+\eta_3 \wedge \xi^2
+\xi^3 \wedge (2 \, \theta_{12}-U_1 \, \theta_{13})
-\theta_{12} \wedge \theta_{13},
\nonumber
\\
\fl
d\theta_{22} &=&
(\eta_4 +U_2\,\theta_3  - U_2 \, \theta_{12}+\theta_{22}+2 \, \theta_{23})\wedge \xi^1
+\eta_7 \wedge \xi^2
-(2\,\eta_1 +\theta_{12}) \wedge \theta_{22}
\nonumber
\\
\fl
&&
+(\eta_6 +(U_1+2) \, \theta_{22}-U_2 \, \theta_{23})\wedge \xi^3
+\theta_2 \wedge (U_2 \, \xi^3 - \xi^1 - \theta_{23}),
\nonumber
\\
\fl
d\theta_{23} &=&
(\eta_3 +\theta_{12}-U_2 \, \theta_{13}+\theta_{23})\wedge \xi^1
+\eta_6 \wedge \xi^2
+(\eta_4 -U_2 \, \theta_{12}+(U_1+1) \, \theta_{23})\wedge \xi^3
\nonumber
\\
\fl
&&
-(\eta_1+\theta_{12}) \wedge \theta_{23}
-\theta_2 \wedge \theta_{12},
\nonumber
\\
\fl
d\eta_1 &=& 0,
\nonumber
\\
\fl
d\eta_2 &=&
\eta_8 \wedge \xi^1
+2\,\eta_2 \wedge (\theta_{12}-\eta_1-\theta_3+U_2 \, \xi^2+ U_1 \, \xi^3)
+\xi^3 \wedge (U_1 \, \theta_{13}-4 \, \theta_{12})
\nonumber
\\
\fl
&&
+\theta_{12} \wedge \theta_{13}
-(\eta_3-2 \, (U_1+U_1^2-U_2) \, \xi^3-2 \, (U_1+1) \, \theta_{12}) \wedge \xi^2
\nonumber
\\
\fl
&&
-(\eta_1+\theta_3) \wedge (\theta_{13}+2 \, (U_1+1) \, \xi^2+3 \, \xi^3),
\nonumber
\\
\fl
d\eta_3 &=&
\eta_1 \wedge (\theta_2-(2 \, U_1+1) \, \xi^1-U_2 \, \xi^3)
-\eta_4 \wedge (2 \, (U_1+1) \, \xi^2+2 \, \xi^3+\theta_{13})
+\theta_{13} \wedge \theta_{23}
\nonumber
\\
\fl
&&
+\eta_3 \wedge (2 \, \theta_{12}-\theta_3+U_2 \, \xi^2+(2 \, U_1+1) \, \xi^3)
-\theta_3 \wedge ((2 \, U_1+1) \, \xi^1+U_2 \, \xi^3)
\nonumber
\\
\fl
&&
+U_2\,\eta_2 \wedge \xi^1
+\theta_2 \wedge (2 \, \theta_{12}-\theta_3 - (3 \, U_1+2 \, U_1^2-2 \, U_2+1) \, \xi^2-U_1 \, \theta_{13})
\nonumber
\\
\fl
&&
-\xi^1 \wedge ((3 \, U_1+2 \, U_1^2-3 \, U_2+1) \, \xi^2 + 2 \, (U_1+U_1^2-U_2) \, \xi^3 + 2 \, U_1 \, \theta_{12}
               +U_1 \, \theta_{13})
\nonumber
\\
\fl
&&
+\xi^2 \wedge ((U_1+1) \, \theta_{23}-U_2 \, (3 \, U_1+2 \, U_1^2-2 \, U_2+1) \, \xi^3-\theta_{22})
\nonumber
\\
\fl
&&
+\xi^3 \wedge (\theta_{23}-2 \, U_2 \, \theta_{12}+U_2 \, U_1 \, \theta_{13}),
\nonumber
\\
\fl
d\eta_4 &=&
\eta_3 \wedge (\theta_2+\xi^1-U_2 \, \xi^3)
+(\eta_4 +U_2 \, \xi^1) \wedge \eta_1
+\eta_4 \wedge (\theta_{12}-U_2 \, \xi^2)
-\eta_6 \wedge \xi^2
\nonumber
\\
\fl
&&
+\theta_2 \wedge ((U_1+1) \, \theta_3+\xi^1-U_2 \, \xi^2-(3 \, U_1+2 \, U_1^2-U_2+1) \, \xi^3
-(U_1+1) \, \theta_{12})
\nonumber
\\
\fl
&&
+\theta_2 \wedge \theta_{13}
+\theta_3 \wedge (U_2 \, (U_1+1) \, \xi^3-(U_1+U_2+1) \, \xi^1)
-\xi^2 \wedge (U_2^2 \, \xi^3+U_1 \, \theta_{22})
\nonumber
\\
\fl
&&
+\xi^1 \wedge (2 \, U_1 \, U_2 \, \xi^2-(3 \, U_1+2 \, U_1^2-3 \, U_2+1) \, \xi^3-(U_1+U_2+1) \, \theta_{12}
+U_2 \, \theta_{13})
\nonumber
\\
\fl
&&
+\xi^1 \wedge \theta_{23}
+U_2\,\xi^3 \wedge ((U_1+1) \, \theta_{12}-U_2\, \theta_{13}),
\nonumber
\\
\fl
d\eta_5 &=&
\eta_1 \wedge (3 \, \eta_5-3 \, \xi^2-\theta_{11})
-2\, \eta_2 \wedge (\theta_0+\theta_1+\xi^2+\xi^3)
+2\,\eta_5 \wedge (\theta_{12}-\theta_3 + U_2 \, \xi^2)
\nonumber
\\
\fl
&&
+2\,U_1\,\eta_5 \wedge \xi^3
+\eta_8 \wedge (\theta_0+\xi^3)
+\eta_9 \wedge \xi^1
-\theta_0 \wedge (\theta_{13}+4 \, (U_1+1) \, \xi^2+4 \, \xi^3)
\nonumber
\\
\fl
&&
+\xi^3 \wedge (U_1 \, \theta_{11}-\theta_{13})
-\theta_1 \wedge (\theta_{13}+3 \, (U_1+1) \, \xi^2+3 \, \xi^3)
-\theta_3 \wedge (\theta_{11}+3 \, \xi^2)
\nonumber
\\
\fl
&&
+\xi^2 \wedge (U_2 \, \theta_{11}-4 \, \theta_{12}-\theta_{13})
-\theta_{11} \wedge \theta_{12},
\nonumber
\\
\fl
d\eta_6 &=&
\eta_6 \wedge (2\, \eta_1+ 2 \, \xi^1+(2 \, U_1+1) \, \xi^3+2 \, \theta_{12})
+\eta_4 \wedge (2 \, \theta_2-2 \, U_1 \, \xi^1-3 \, U_2 \, \xi^3-\theta_{23})
\nonumber
\\
\fl
&&
+\theta_2 \wedge (U_2 \, \theta_{12}-(3 \, U_1+2 \, U_1^2-2 \, U_2+1) \, \xi^1-U_2 \, \xi^3-(U_1+1) \, \theta_{23})
+\eta_{10} \wedge \xi^2
\nonumber
\\
\fl
&&
+\xi^1 \wedge (2 \, U_2 \, \theta_{12}-U_2 \, (U_1+2 \, U_1^2-2 \, U_2+1) \, \xi^3-U_2^2 \, \theta_{13}-\theta_{22}
 +U_2 \, \theta_{23})
\nonumber
\\
\fl
&&
+\xi^3 \wedge (U_2 \, (U_1+1) \, \theta_{23}-U_2^2 \, \theta_{12}-U_1 \, \theta_{22})
-\theta_{12} \wedge \theta_{22}
-U_2\,\eta_3 \wedge \xi^1,
\nonumber
\\
\fl
d \eta_7 &=&
\eta_7  \wedge (3\,\eta_1+2 \, \xi^1+(3+2 \, U_1) \, \xi^3+2 \, \theta_{12})
-\eta_4 \wedge (\theta_{22}+3 \, U_2 \, \xi^1)
+\eta_{10} \wedge \xi^3
\nonumber
\\
\fl
&&
+\eta_6 \wedge (2 \, \theta_2+2 \, \xi^1-2 \, U_2 \, \xi^3)
+\theta_2 \wedge (U_2 \, \xi^1-U_2^2 \, \xi^3-(U_1+2) \, \theta_{22}+U_2 \, \theta_{23})
\nonumber
\\
\fl
&&
+\eta_{11} \wedge \xi^2+\xi^1 \wedge (U_2^2\,(\theta_3 -\xi^3-\, \theta_{12})-(2 \, U_1-U_2+3) \, \theta_{22}+4 \, U_2 \, \theta_{23})
\nonumber
\\
\fl
&&
+\theta_{22} \wedge \theta_{23}
+U_2\,\xi^3 \wedge ((U_1+3) \, \theta_{22}-U_2 \, \theta_{23}),
\label{d_theta_eqns}
\end{eqnarray}
and
\begin{eqnarray}
\fl
d\omega_0 &=&
\omega_0 \wedge (\omega_3 -\theta_{12}   +V_1 \, \xi^1+V_2 \, \xi^2+V_3 \, \xi^3)
+\xi^1 \wedge \omega_1
+\xi^2 \wedge \omega_2
+\xi^3 \wedge \omega_3,
\nonumber
\\
\fl
d\omega_1 &=&
\omega_0 \wedge \theta_{13}
+\omega_1 \wedge (\omega_3-\eta_1-\theta_3)
+\eta_{12} \wedge \xi^1
+\eta_{13} \wedge \xi^2
+\eta_{14} \wedge \xi^3,
\nonumber
\\
\fl
d\omega_2 &=&
\omega_2 \wedge (\eta_1+(V_1+1) \, \xi^1+\omega_3)
+\omega_3 \wedge (\theta_2+\xi^1)
+\eta_{15} \wedge \xi^2
+\eta_{16} \wedge \xi^3
\nonumber
\\
\fl
&&
+(\eta_{13} -(U_1+1)\,\omega_0-(U_2+V_2)\,\omega_1)\wedge \xi^1,
\nonumber
\\
\fl
d\omega_3 &=&
\omega_3 \wedge ((V_1+1) \, \xi^1+(U_2+V_2) \, \xi^2+(U_1+V_3) \, \xi^3)
-\omega_0 \wedge (\xi^1+(U_1+1) \, \xi^3)
\nonumber
\\
\fl
&&
-\omega_1 \wedge ((U_1+V_3) \, \xi^1+(U_2+V_2) \, \xi^3)
+\eta_{14} \wedge \xi^1
+(\eta_{16} -(U_1+V_3+1)\,\omega_2) \wedge \xi^2
\nonumber
\\
\fl
&&
+(\eta_{13} + (V_1+1)\, \theta_2 - (U_1+V_3)\,\theta_3 -\theta_{12}-V_4 \, \theta_{13})\wedge \xi^3,
\nonumber
\\
\fl
d\eta_{12} &=&
\eta_{17} \wedge \xi^1
+(V_1+1) \,(\omega_1 \wedge \omega_3+\eta_{13}\wedge \xi^2+\eta_{14} \wedge \xi^3)
+\xi^3 \wedge (\eta_2-(V_1+2)\,\theta_{13})
\nonumber
\\
\fl
&&
+\omega_1 \wedge (\eta_{14}+\xi^2+V_1 \, \xi^3)
+\omega_0 \wedge (\omega_1-2 \, (U_1+1) \, \xi^2-2 \, \xi^3+V_1 \, \theta_{13}+\eta_2)
\nonumber
\\
\fl
&&
+\eta_{12} \wedge (\omega_3-2\,\eta_1-2 \, \theta_3+(2 \, U_2+V_2) \, \xi^2+(V_3+2 \, U_1) \, \xi^3+\theta_{12})
\nonumber
\\
\fl
&&
+\xi^2 \wedge ((2 \, V_4+U_1 \, (V_1+2)+V_1-V_3+3) \, \xi^3 -V_4\,\eta_2-(U_1+V_3-V_4) \, \theta_{13}),
\nonumber
\\
\fl
d\eta_{13} &=&
\omega_0 \wedge (\eta_3-2 \, (U_1+1) \, (\xi^1+2 \, U_2 \, \xi^2)-2 \, U_2 \, \xi^3+V_2 \, \theta_{13})
+\omega_1 \wedge (\eta_{16}+\theta_{23}+\xi^1)
\nonumber
\\
\fl
&&
+\omega_1 \wedge ((U_2+V_2) \, \omega_3-(U_1+V_3+1) \, \omega_2
-(V_4\,(U_1+1)-2 \, (U_2+V_2)) \, \xi^3)
\nonumber
\\
\fl
&&
+2 \, \eta_{16} \wedge \xi^2
-\omega_2 \wedge (\theta_{13}+2 \, (2 \, U_1+V_3+2) \, \xi^2+\xi^3)
-\omega_3 \wedge (\eta_{13}-(U_1+2) \, \xi^3)
\nonumber
\\
\fl
&&
+\eta_{13} \wedge (\theta_{12}-\theta_3+(V_1+2) \, \xi^1+(3 \, U_2+2 \, V_2) \, \xi^2+(2 \, U_1+V_3+1) \, \xi^3)
\nonumber
\\
\fl
&&
+\eta_{14} \wedge (\theta_2+\xi^1+(U_2+V_2) \, \xi^3)
+\theta_2 \wedge ((U_1+V_3-2) \, \xi^2-V_1 \, \xi^3)
\nonumber
\\
\fl
&&
+\theta_3 \wedge (2 \, (U_2+V_2) \, \xi^2+\xi^3)
+\xi^3 \wedge (2 \, \theta_{12}-(U_2+V_2) \, \theta_{13}-(V_1+1) \, \theta_{23})
\nonumber
\\
\fl
&&
+(V_4\,\eta_2 
+(U_1+V_3-V_4) \, \theta_{13}
)\wedge \xi^1
+\xi^2 \wedge ((V_1+1) \, \theta_{22}+2 \, (U_2+V_2) \, \theta_{12})
\nonumber
\\
\fl
&&
+\xi^1 \wedge ((U_2+V_2) \, \eta_{12}+(2 \, V_4\,(U_1+1)+3 \, U_1+U_2 \, (V_1-1) -2 \, V_2+V_3+2) \, \xi^2
)
\nonumber
\\
\fl
&&
+(2 \, V_4+V_1\,(U_1-1)-V_3-2) \,\xi^1 \wedge  \xi^3
+\xi^2 \wedge (V_5 \, \theta_{13}-(U_1+V_3+2) \, \theta_{23})
\nonumber
\\
\fl
&&
+(V_5+(U_1\,(2\,U_2+3 \, V_2+2 \, V_4)-U_2\,(V_3+4)-2 \, V_2+2 \, V_4) \, \xi^2 \wedge \xi^3,
\nonumber
\\
\fl
d\eta_{14} &=&
\omega_0 \wedge (\eta_1+\theta_3-2 \, \xi^1-2 \, U_2 \, \xi^2-(2 \, U_1+1) \, \xi^3
-2 \, \theta_{12}+(U_1+V_3) \, \theta_{13})
\nonumber
\\
\fl
&&
+\omega_1 \wedge ((V_1+1) \, \theta_2-(U_1+V_3) \, \theta_3
-(V_4\,(U_1+1)-2 \,(U_2+ V_2)) \, \xi^2
-(V_4-1) \, \xi^3)
\nonumber
\\
\fl
&&
+\omega_1 \wedge (\eta_{13}-(U_1+1)\,\omega_0+(U_1+V_3) \, \omega_3-V_4 \, \theta_{13}+V_1 \, \xi^1)
-(V_1+1) \,\xi^2 \wedge \theta_{23}
\nonumber
\\
\fl
&&
-(\eta_2 -(U_1+V_3)\,\eta_{12}) \, \wedge \xi^1
-(\omega_2  -(U_1+V_3)\,\eta_{13}+V_1\,\theta_2)\wedge \xi^2
+\eta_1 \wedge (\eta_{14}-\xi^3)
\nonumber
\\
\fl
&&
+\eta_{14} \wedge (\theta_{12}
-\theta_3+(V_1+2) \, \xi^1+(2 \, U_2+V_2) \, \xi^2+(3 \, U_1+2 \, V_3+1) \, \xi^3)
\nonumber
\\
\fl
&&
-\xi^1 \wedge ((V_1+2) \, \theta_{13}+(2 \,(U_1+ V_1)+5) \, \xi^2)
-\xi^3 \wedge (2 \, \theta_{12}+(U_1+V_3+1) \, \theta_{13})
\nonumber
\\
\fl
&&
+\xi^2 \wedge (2 \, \theta_{12}
+(U_1 \,(U_1+V_3 +5) - U_2\,(V_1-1)+2 \, V_3+3) \, \xi^3
-(U_2+V_2) \, \theta_{13})
\nonumber
\\
\fl
&&
-\omega_3 \wedge (\eta_{14}+(2+U_1) \, \xi^2+\xi^3+\theta_{13})
+\theta_3 \wedge (\xi^2-\xi^3),
\nonumber
\\
\fl
d\eta_{15} &=&
(U_2+V_2)\,(\eta_{13}+2\,\theta_3 - (U_1+1) \,\omega_0 -(U_2+V_2)\,+\omega_1) \wedge \xi^1
+\eta_6 \wedge (V_4 \, \xi^2-\xi^3)
\nonumber
\\
\fl
&&
+\omega_2 \wedge (\eta_{16}
+(V_2+U_2) \, \omega_3
-(U_1+V_3+1) \, \theta_2
-(V_4\,(U_1+1)-V_2) \, \xi^3)
\nonumber
\\
\fl
&&
+((V_1 +1)\, (U_2+V_2)-3 \, (U_1+V_3)-2) \,\omega_2 \wedge  \xi^1
-\eta_7 \wedge \xi^2
\nonumber
\\
\fl
&&
-\omega_3 \wedge (\eta_{15}+\theta_{22}-(U_2+V_2) \, \theta_2-V_2 \, \xi^1)
+\eta_{16} \wedge (2 \, \theta_2+4 \, \xi^1+(U_2+V_2) \, \xi^3)
\nonumber
\\
\fl
&&
+\eta_{15} \wedge (2\,\eta_1+\theta_{12}+(V_1+2) \, \xi^1+(3 \, U_2+2 \, V_2) \, \xi^2+(2 \, U_1+V_3+2) \, \xi^3)
\nonumber
\\
\fl
&&
+\theta_2 \wedge ((U_1+V_3-2) \, \xi^1
+(2 \, U_2 \, V_4+3 \, V_5) \, \xi^2
+(2 \,V_4\,(U_1 +1)-5 \, (U_2+V_2)) \, \xi^3 )
\nonumber
\\
\fl
&&
+\xi^1 \wedge (
2 \, (U_2+V_2) \, \theta_{12}+V_5 \, \theta_{13}-(V_1+1) \, \theta_{22}-(U_1+V_3+2) \, \theta_{23}
)
\nonumber
\\
\fl
&&
-(2 \, U_1\,(U_2 \, V_4+\, V_5)
-U_2\,(7 \, U_2-5 \, V_2-V_4)-V_5) \,\xi^1 \wedge  \xi^2
\nonumber
\\
\fl
&&
+(U_1\,(7 \, U_2+5 \, V_2+4 \,  V_4)-2 \, U_2-4 \, (V_2-V_4)+V_5) \,\xi^1 \wedge \xi^3
\nonumber
\\
\fl
&&
+\xi^2 \wedge ( (3 \, V_5+2 \, U_2 \, V_4) \, (\theta_{23}+U_2\,\xi^3)
-(4 \, U_2+3 \, V_2+V_4) \, \theta_{22})
\nonumber
\\
\fl
&&
-\xi^3 \wedge ((U_1+V_3) \, \theta_{22}+(4 \, U_2+3 \, V_2) \, \theta_{23}),
\nonumber
\\
\fl
d \eta_{16} &=&
((U_1+1) \, \omega_{0} +(U_2+V_2) \, \omega_{1} )\wedge (\omega_{2}-\theta_2- (U_1+V_3+2) \, \xi^1)
-\eta_1 \wedge \eta_{16}
+\theta_2 \wedge \theta_{12}
\nonumber
\\
\fl
&&
+\omega_{2} \wedge ((V_1+1) \, \theta_2
-(U_1+V_3) \, \theta_3
+((U_1+V_3) \, (V_1+1)+V_1) \, \xi^1
-\theta_{12}-V_4 \, \theta_{13})
\nonumber
\\
\fl
&&
+\omega_{2} \wedge (\eta_{13}+(U_1+V_3) \, \omega_{3}
-(V_4\,(U_1+1)-V_2) \, \xi^2
-(5 \, U_1+3 \, V_3+V_4+3) \, \xi^3)
\nonumber
\\
\fl
&&
-\omega_{3} \wedge (\eta_{16}-(U_1+V_3+1) \, \theta_2-(V_3-1) \, \xi^1+\theta_{23})
+\eta_{13} \wedge (\theta_2+(U_1+V_3+2) \, \xi^1)
\nonumber
\\
\fl
&&
-\xi^2 \wedge ((U_1+V_3+1) \,\eta_{15} 
-(V_5\,(U_1+1)+U_2\,(2 \,V_4\,( U_1+1)-3 \,(U_2+ V_2))) \, \xi^3
)
\nonumber
\\
\fl
&&
+\xi^2 \wedge (\eta_6
-(U_1+V_3) \, \theta_{22}-(4 \, U_2+3 \, V_2) \, \theta_{23}
)
+\xi^1  \wedge (3 \, \theta_{12}+(U_1+V_3-1) \, \theta_3)
\nonumber
\\
\fl
&&
+\xi^3 \wedge (\eta_4+V_4\,\eta_3
- V_4 \, (U_1+1) \,\theta_3
+V_2 \, \theta_{12}-U_2 \, V_4 \, \theta_{13}-2 \, (U_1+V_3+1) \, \theta_{23})
\nonumber
\\
\fl
&&
+\eta_{16} \wedge ((V_1+2) \, \xi^1+(2 \, U_2+V_2) \, \xi^2+(3 \, U_1+2 \, V_3+4) \, \xi^3+\theta_{12})
\nonumber
\\
\fl
&&
+\theta_2 \wedge ((U_1+V_3) \, \theta_3+\xi^1
+(2 \, V_4\,(U_1+1)-5 \, (U_2+V_2)) \, \xi^2
+(2 \, U_1+V_3) \, \xi^3)
\nonumber
\\
\fl
&&
+\xi^1 \wedge ((U_1\,(2\,(V_2+V_4)+5\,U_2)+U_2\,(V_3+2)-2\,(V_2+V_4)) \, \xi^2
-(V_1+1) \, \theta_{23}
)
\nonumber
\\
\fl
&&
+\xi^1 \wedge (U_1\,(U_1+V_3+2  \, V_4+8)-2\,(V_2+V_4)+3 \, V_3+5) \, \xi^3
\nonumber
\\
\fl
&&
+(V_4\,\theta_2-(U_2+V_2-V_4)\,\xi^1)\, \wedge  \theta_{13}.
\label{d_omega_eqns}
\end{eqnarray}
Eqns. (\ref{d_theta_eqns}) are the involutive structure equations for the symmetry pseudo-group of
Eq. (\ref{universal_hierarchy_eq}).
The invariants $U_1$, $U_2$, $V_1$, $V_2$, $V_4$ in (\ref{d_theta_eqns}), (\ref{d_omega_eqns})   have the 
follownig expressions
\begin{eqnarray}
\fl
U_1 &=& \frac{u_x\,(u_y^2 \, u_{txy}+S_2^2+u_x \, S_1) - u_y \,(u_{xy} \, S_2-u_y \, u_{tx} \, u_{xy})}
{u_x\,S_2^2},
\nonumber
\\
\fl
U_2 &=& S_1\,
\left(
\frac{u_y}{u_xS_2^3}\,(u_y u_{tx} - u_x u_{xy}) 
+\frac{u_y^3}{u_xS_2^4}\,(u_x u_{txx} - u_{tx} u_{xx})
+\frac{S_1 u_x^2}{S_2^4}
+\frac{u_x}{S_2^2}\,(2\,U_1+1)
\right),
\nonumber
\\
\fl
V_1 &=& \frac{u_x \, u_y \, (v_{ty} \, S_2 - v_y \, S_{2t})}{v_y\,(u_x \, u_y \, S_{2t}-u_y \, u_{tx} \, S_2-S_2^2)},
\nonumber
\\
\fl
V_2 &=& 
\frac{ u_y^2 \, (v_x-v_y) \, v_{ty}}{v_y^2 S_2}
+\frac{(V_1+1) \, (u_x\,v_y-u_y \, v_x) \, S_0}{S_2}-U_1,
\nonumber
\\
\fl
V_4 &=& (u_y \, v_x-u_x\,v_y) \, S_0\,S_2^{-1},
\nonumber
\end{eqnarray}
where
\begin{eqnarray}
S_0 &=& (u_y \, u_x \, S_{2t}-u_y \, u_{tx} \, S_2-S_2^2)\,(u_x v_y S_2)^{-1},
\nonumber
\\
S_1 &=& u_x \, u_y \, u_{tty}-u_y^2 \, u_{ttx}+u_y \, u_{tx} \, u_{ty}-u_x \, u_{ty}^2,
\nonumber
\\
S_2 &=&  u_x\,u_{ty}-u_y\,u_{tx},
\nonumber
\end{eqnarray}
while the formulas for $V_3$ and $V_5$ are too big to write them in full here.
The differentials of the invariants satisfy the equations
\begin{eqnarray}
\fl
dU_1 &=& U_2\,\theta_{13}-\eta_3-\theta_2+(U_1+1)\,\theta_1-\xi^1
-(2\,U_1^2+3\,U_1-2\,U_2+1)\,\xi^3-2\,(U_1+1)\,\theta_{12},
\nonumber
\\
\fl
dU_2 &=&
 -U_1\,\theta_2-(2\,U_1+1)\,\xi^1-\theta_{23}-\eta_4
-U_2\,(\eta_1+\theta_{12}+\xi^3),
\nonumber
\\
\fl
d V_1 &=& \omega_0+(U_1+V_3) \, \omega_1
+(V_1+1) \, (\eta_1-\omega_3+\theta_3-\theta_{12})
-\eta_{14}
 -\theta_{13}
-(V_1^2+3 \, V_1+2) \, \xi^1
\nonumber
\\
\fl
&&
- (U_1+(2\,U_2+V_3)\,(V_1+1)+2) \, \xi^2
-(V_1+1)(2\,U_1+V_2+1) \, \xi^3,
\nonumber
\\
\fl
d V_2  &=& 
(U_1+V_3+1) \, \omega_2-V_2 \, \eta_1-(U_2+V_2) \, \omega_3-\eta_{16}+\eta_4
-V_3 \, \theta_2-V_2 \, \theta_{12}
\nonumber
\\
\fl
&&
-(U_1\,(2\,U_2+V_3-2\,V_4)+2\,U_2\,(V_2+1)+V_3\,(V_2+2)-V_4)\,\xi^2
\nonumber
\\
\fl
&&
-((U_1+V_2)\,(V_1+2)+V_1)\,\xi^1
-(V_2\,(V_2+3\,U_1+1)+2\,(U_2-U_1)-V_4)\,\xi^3,
\nonumber
\\
\fl
d V_3  &=& (U_1+1) \, \omega_0
+\eta_3
-\eta_{13}
+(U_2+V_2) \, \omega_1
-(U_1+V_3) \, \omega_3
-V_1 \, \theta_2
+(V_3-1) \, \theta_3
\nonumber
\\
\fl
&&
+(U_1-V_3+2) \, \theta_{12}+(V_4-U_2) \, \theta_{13}
-((U_2+V_3)\,(V_1+2)+V_2+1)\,\xi^1
\nonumber
\\
\fl
&&
-(U_1\,(2\,U_2+2\,V_3-V_4)+U_2\,(V_2+2)+V_3\,(V_2+3)-V_4)\, \xi^3
\nonumber
\\
\fl
&&
-(U_2+V_3)\,(2\,U_2+V_3)\,\xi^2,
\nonumber
\\
\fl
d V_4 &=& 
\theta_2-\omega_2-V_4 \, (\eta_1+ \omega_3)
-(V_4\,(V_4+1)+U_1+V_3-1) \, \xi^1
+(V_5-V_4\,(U_2+V_2)) \, \xi^2
\nonumber
\\
\fl
&&
-(V_4\,(U_1+V_3+1)+U_2+V_2) \, \xi^3,
\nonumber
\\
\fl
d V_5 &=& 
\eta_{15}
+\theta_{22}
-(U_2+V_2) \, \omega_2
-V_4 \, \theta_{23}
-V_5 \, (\omega_3+2 \,\eta_1+\theta_{12})
+(V_4+2 \, (U_2+V_2)) \, \theta_2
\nonumber
\\
\fl
&&
+(5 \, U_2+4 \, V_2-U_1 \, V_4-V_5\,(V_1+2)) \, \xi^1
-(U_2 \, V_4+V_5\,(2\,U_1+V_3+2)) \, \xi^3
\nonumber
\\
\fl
&&
-V_5 \, (2 \, U_2+V_2) \, \xi^2.
\nonumber
\end{eqnarray}
In what follows we need explicit formulas only for the {\sc mc} forms 
\begin{eqnarray}
\fl
\theta_0 &=& \frac{u_y\, u_{tx}-u_x\,u_{ty}}{u_x^2}\, (du - u_t\,dt- u_x\,dx- u_y\,dy),
\nonumber
\\
\fl
\theta_1 &=& -\frac{S_2^2}{u_y\,S_1}\, (du_t - u_{tt}\,dt- u_{tx}\,dx- u_{ty}\,dy)
+\frac{u_{tx}\, S_2}{S_1}\, \theta_0,
\nonumber
\\
\fl
\theta_2 &=& \frac{S_1}{u_y\,S_2^2}\, (u_x\, d u_y - u_y\, d u_x -S_2\, dt -(u_y\, u_{xx}-u_x\, u_{xy}) dx
+(u_y\, {u_xy} + u_x\, S_2)  dy),
\nonumber
\\
\fl
\theta_{12} &=& \frac{1}{S_2}\, 
\left( d S_2 +u_{tx} du_y + u_{ty} du_x
-\frac{u_x S_1 + u_y u_{tx} S_2 + S_2^2}{u_x\,u_y}\, dt 
- (u_x u_{txy}-u_y u_{txx})\, dx
\right.
\nonumber
\\
\fl
&&
\qquad\qquad
\left.
+\frac{u_y^2 u_{txy}+u_x S_1 +u_y u_{tx} S_2 +S_2^2}{u_y}\, dy\right)
-\frac{S_2}{S_1\, u_{ty}}\, \theta_2,
\nonumber
\\
\fl
\xi^1 &=& \frac{S_1}{u_y\,S_2}\, dt,
\qquad
\xi^2 = \frac{S_2^3}{u_y^2\,S_1}\, dx,
\qquad
\xi^3 = -\frac{S_2}{u_y^2}\, (u_x dx + u_y dy),
\nonumber
\\
\fl
\omega_0 &=& -\frac{S_2}{u_y\,v_y}\, (dv - v_t\,dt- v_x\,dx- v_y\,dy),
\nonumber
\\
\fl
\omega_1 &=& -\frac{S_2}{v_y\,S_1}\, (dv_t - v_{tt}\,dt- v_{tx}\,dx- v_{ty}\,dy) 
- \frac{S_2\,u_{ty}}{S_1}\, \omega_0,
\nonumber
\\
\fl
\omega_3  &=& \frac{1}{v_y}\, 
(dv_y - v_{ty}\,dt- v_{xy}\,dx- (u_y v_{tx}-u_x v_{ty} + v_y u_{tx}-v_x u_{ty})\,dy). 
\label{MCforms}
\end{eqnarray}

\section{Contact integrable extensions}

For applying \'Elie Cartan's structure theory of Lie pseudo-groups to the problem of finding coverings of 
{\sc pde}s we use the notion of integrable extension. It was introduced in \cite{BryantGriffiths1995} for the 
case of {\sc pde}s with two independent variables and finite-di\-men\-si\-o\-nal coverings. The generalization of 
the definition to the case of infinite-di\-men\-si\-o\-nal coverings of {\sc pde}s with more than two independent 
variables was proposed in \cite{Morozov2009b}.  In contrast to \cite{WahlquistEstabrook1975,BryantGriffiths1995}, 
the starting point of our definition is the set of Maurer--Cartan forms of the symmetry pseudo-group of a given 
{\sc pde}, and all the constructions are carried out in terms of in\-va\-ri\-ants of the pseudo-group. Therefore, the 
effectiveness of our method increases when it is applied to equations with large symmetry pseudo-groups.

\hspace{20pt}
Let $\mathfrak{G}$  be a Lie pseudo-group on a manifold $M$. Let  $\omega^1$, ... , $\omega^m$, 
$m=\mathrm{dim}\,M$,  be its Maurer--Cartan forms with the compatible and involutive structure equations 
(\ref{SE}), (\ref{dUs}).  Consider the system of exterior differential equations
\begin{eqnarray}
\fl
d\zeta^q &=& 
D^q_{\rho r} \, \mu^\rho \wedge \zeta^r 
+ 
E^q_{r s} \, \zeta^r \wedge \zeta^s 
+
F^q_{r \beta} \, \zeta^r \wedge \pi^\beta
+
G^q_{r j} \, \zeta^r \wedge \omega^j
+
H^q_{\beta j} \, \pi^\beta \wedge \omega^j
\nonumber
\\
\fl
&&+
I^q_{j k} \, \omega^j \wedge \omega^k,
\label{extra_SE} 
\\
\fl
d V^\epsilon &=&  J^\epsilon_j \,\omega^j 
+  K^\epsilon_q \, \zeta^q,
\label{dVs}
\end{eqnarray}
for unknown 1-forms $\zeta^q$, $q \in \{1,...,Q\}$, $\mu^\rho$, $\rho \in \{1,...,R\}$, and unknown functions 
$V^\epsilon$, $\epsilon \in \{1,...,S\}$ with some $Q, R, S \in \mathbb{N}$. The coefficients $D^\kappa_{\rho r}$, 
..., $K^\epsilon_q$ in equations  (\ref{extra_SE}), (\ref{dVs}) are supposed to be functions of  
$U^\lambda$ and $V^\kappa$.

\vskip 10 pt
\noindent
{\sc definition} 1.
The system (\ref{extra_SE}), (\ref{dVs}) is called an  {\it integrable extension} of the system 
(\ref{SE}), (\ref{dUs}), if equations  (\ref{extra_SE}), (\ref{dVs}), 
(\ref{SE}), (\ref{dUs}) together meet the involutivity conditions and the compatibility conditions 
\begin{equation}
d(d\zeta^q) \equiv 0,
\qquad 
d(d V^\epsilon) \equiv 0.
\label{compatibility_condition}
\end{equation}

\vskip 10 pt

Equations  (\ref{compatibility_condition}) give an over-determined system of {\sc pde}s for the coefficients 
$D^\kappa_{\rho r}$, ..., $K^\epsilon_q$ in equations  (\ref{extra_SE}), (\ref{dVs}). Suppose this system is satisfied. Then we apply the third and the second inverse fundamental Lie's theorems in Cartan's form, \cite[\S\S 16--24]{Cartan1}, \cite{Cartan4}, 
\cite[\S\S 16, 19, 20, 25, 26]{Vasilieva1972}, \cite[\S\S 14.1--14.3]{Stormark2000}. The third inverse fundamental theorem ensures the existence of the forms $\zeta^q$ and the functions $V^\epsilon$, the solutions to equations (\ref{extra_SE}), (\ref{dVs}). In accordance with the second inverse fundamental theorem, the forms $\zeta^q$,  $\omega^i$  are Maurer--Cartan forms for a  Lie pseudo-group  $\mathfrak{H}$  acting on  $M \times \mathbb{R}^Q$.

\vskip 10 pt
\noindent
{\sc definition} 2.
The integrable extension (\ref{extra_SE}), (\ref{dVs}) is called {\it trivial}, if there exists a change of variables on the manifold of action of the pseudo-group  $\mathfrak{H}$ such that in the new coordinates the coefficients $F^q_{r \beta}$, $G^q_{r j}$, $H^q_{\beta j}$, $I^q_{j k}$ and $J^\epsilon_j$ are identically equal to zero, while the coefficients  $D^q_{\rho r}$, 
$E^q_{r s}$ and $K^\epsilon_q$ are independent of  $U^\lambda$.  Otherwise, the integrable extension is called  {\it nontrivial}.

\vskip 10 pt

Let  $\theta^\alpha_I$ and $\xi^j$ be a set of Maurer--Cartan forms of a symmetry pseudo-group $\mathfrak{Lie}(\EuScript{E})$ of a {\sc pde}  $\EuScript{E}$ such that $\xi^i$ are horizontal forms, that is, $\xi^1 \wedge ... \wedge \xi^n \not = 0$ on each solution of $\EuScript{E}$, while   $\theta_I^\alpha$ are contact forms, that is, they are equal to 0 on each solution.

\vskip 10 pt
\noindent
{\sc definition} 3. 
Nontrivial integrable extension of the structure equations for the pseudo-group $\mathfrak{Lie}(\EuScript{E})$  of the form
\begin{equation}
d \omega^q =\Pi^q_r \wedge \omega^r + \xi^j \wedge \Omega^q_j, 
\label{contact_ie}
\end{equation}
$q,r \in \{1, \dots , N\}$, $N \ge 1$, is called a 
{\it contact integrable extension}, if the following conditions are satisfied:

\begin{enumerate}

\item[({\it i})]
$\Omega^q_j \in \langle \theta^{\alpha}_I, \, \omega^r_i\rangle_{\tt lin}$ for some additional 1-forms  $\omega^r_i$;

\item[({\it ii})]
$\Omega^q_j \not \in \langle  \omega^r_i\rangle_{\tt lin}$ for some $q$ and $j$;

\item[({\it iii})]
$\Omega^q_j \not \in \langle  \theta^{\alpha}_I \rangle_{\tt lin}$  for some $q$ and $j$;

\item[({\it iv})] 
$\Pi^q_r \in \langle \theta^{\alpha}_I, \,\xi^j, \, \omega^r, \, \omega^r_i\rangle_{\tt lin}$.

\item[({\it v})]
The coefficients of expansions of the forms $\Omega^q_j$ with respect to  $\{\theta^{\alpha}_I, \, \omega^r_i\}$
and the forms $\Pi^q_r$ with respect ot  $\{\theta^{\alpha}_I, \,\xi^j, \, \omega^r, \, \omega^r_i\}$
depend either on the invariants of the pseudo-group $\mathfrak{Lie}(\EuScript{E})$ alone, or they depend also on a set of some additional functions   $W_\rho$, $\rho \in \{1, \dots, \Lambda\}$, 
$\Lambda \ge 1$. In the latter case, there exist functions $P^{I \rho}_\alpha$, $Q^\rho_q$, $R_q^{j\rho}$  and $S_j^\rho$ such that
\begin{equation}
dW_\rho = P^{I}_{\rho \alpha}\,\theta^\alpha_I+Q_{\rho q}\,\omega^q+R_{\rho q}^{j}\,\omega^q_j+S_{\rho j}\,\xi^j,
\label{dW_in_the_definition}
\end{equation}
and the set of equations  (\ref{dW_in_the_definition}) satisfies the compatibility conditions
\begin{equation}
d(dW_\rho)  
= d \left(
P^{I}_{\rho \alpha}\,\theta^\alpha_I+Q_{\rho q}\,\omega^q+R_{\rho q}^{j}\,\omega^q_j+S_{\rho j}\,\xi^j\right)
\equiv  0.
\label{compatibility_for_dW}
\end{equation}

\end{enumerate}

\vskip 10 pt

\noindent
{\sc example} 2.  Eqns.  (\ref{d_omega_eqns}) are a {\sc cie} for Eqns. (\ref{d_theta_eqns}) with
the additional forms $\eta_{12}$, ... , $\eta_{17}$ and the additional invariants $V_1$, ... , $V_5$.

\vskip 15 pt

\section{B\"acklund auto-transformation for the tangent covering of the universal hierarchy equation}

We apply Definition 3 to the structure equations (\ref{d_theta_eqns}), (\ref{d_omega_eqns}). We restrict our 
analysis to {\sc cie}s of the form 
\begin{eqnarray}
\fl
d \omega_4 
&=& 
\sum \limits_{k=0}^4
\left(
\sum \limits_{i=0}^3 A_{ik} \,\theta_i 
+ \sum {}^{*} B_{ijk}\,\theta_{ij}
+ \sum \limits_{s=1}^{7} C_{sk}\,\eta_s
+ \sum \limits_{s=12}^{16} C_{sk}\,\eta_s 
+ \sum \limits_{j=1}^3 D_{jk}\,\xi^j 
\right) \wedge \omega_k 
\nonumber
\\
\fl
&&
+
\sum \limits_{k=0}^4 E_k\,\omega_k \wedge \omega_5
+
\sum \limits_{k=1}^3 \left(
\sum \limits_{i=0}^3 F_{ik}\,\theta_i 
+ \sum {}^{*} G_{ijk}\,\theta_{ij}
+ H_k\,\omega_5
\right) \wedge \xi^k
\nonumber
\\
\fl
&&
+\sum_{k=0}^3 M_k\, \omega_k \wedge \omega_4
\label{ie_main}
\end{eqnarray}
with one additional form $\omega_5$ mentioned in the part (i) of Definition 3.
 In (\ref{ie_main}),  
$\sum {}^{*}$ means summation for all $i,j \in \mathbb{N}$ such that $1\le i \le j \le 3$, $(i,j)\not = (3,3)$. 
These equations together with equations (\ref{d_theta_eqns}), (\ref{d_omega_eqns}) satisfy the requirement of 
involutivity. We assume that the coefficients of (\ref{ie_main}) depend 
on the invariants $U_1$, $U_2$, $V_1$, ... , $V_5$.

\vskip 10 pt
\noindent
{\sc remark} 3.   We consider (\ref{d_omega_eqns}),  (\ref{ie_main}) together as a single {\sc cie} for 
(\ref{d_theta_eqns}). This defines the form of the r.h.s. of (\ref{ie_main}).

\vskip 10 pt

Definition 3 
gives an over-determined system of {\sc pde}s for the coefficients of (\ref{ie_main}). 
The analysis of this system gives a  {\sc cie}. Then, since the {\sc mc} forms included in Eqns.
(\ref{d_theta_eqns}), (\ref{d_omega_eqns})  are known explicitly, we use the third inverse fundamental Lie's  
theorem and find form $\omega_4$ by means of integration. We obtain

\vskip 10 pt
\noindent
{\sc theorem}. 
{\it   
The structure equations (\ref{d_theta_eqns}), (\ref{d_omega_eqns}) of the contact symmetry pseudo-group of the 
tangent covering of Eq. (\ref{universal_hierarchy_eq})
 have the {\sc cie} 
\begin{eqnarray}
\fl
d\omega_4 &=& 
\left(\eta_1-\omega_3+\theta_{12}-V_2\,\xi^2-V_3\,\xi^3\right) \wedge \omega_4 
+\omega_5 \wedge \xi^1 + \omega_1 \wedge \theta_0  - \omega_0 \wedge \theta_1
\nonumber
\\
\fl
&&
+(\omega_3 -\omega_0 -(U_1+V_3)\,\theta_0 -V_4 \,\theta_1) \wedge \xi^2
+( \omega_1 + \theta_1-(V_1+1)\,\theta_0) \wedge \xi^3.
\nonumber
\end{eqnarray}
Each solution of this equation is contact-equivalent to the form
\begin{eqnarray}
\fl
\omega_4 &=& \frac{(u_y u_{tx}-u_x u_{ty})^3}{u_y^2\,v_y\,S_1}\,
(dw-w_t\,dt-(v_y+u_x v_t-u_{tx} v)\,dx-(u_yv_t-u_{ty}v)\,dy)
\nonumber
\\
\fl
&&
-\frac{v\,(u_y u_{tx}-u_x u_{ty})}{u_y\,v_y}\, \theta_1
-\frac{(u_y v_t - u_{ty} v)\,(u_y u_{tx}-u_x u_{ty})^2}{u_y\,v_y\,S_1}\, \theta_0.
\nonumber
\end{eqnarray}
}

\vskip 10 pt
The form $\omega_4$ is equal to zero on solutions of (\ref{universal_hierarchy_eq}) whenever $w$ solves the system
\begin{equation}
\left\{
\begin{array}{lcl}
w_x &=& v_y+u_x  v_t-u_{tx} v,
\\
w_y &=& u_yv_t-u_{ty}v.
\end{array}
\right.
\label{direct_BT}
\end{equation}
This system is compatible on solutions to (\ref{linearized_universal_hierarchy_eq}). Thus it defines a B\"acklund 
transformation from (\ref{linearized_universal_hierarchy_eq}) to a certain {\sc pde}. To get this {\sc pde}, we solve
(\ref{direct_BT}) for $v_t$ and $v_y$:
\begin{equation}
\left\{
\begin{array}{lcl}
v_t &=& (w_y+u_{ty}\,v)\,u_y^{-1},
\\
v_y &=&  ((u_y u_{tx}- u_x u_{ty})\,v+u_y w_x - u_x w_y)\,u_y^{-1}.
\end{array}
\right.
\label{inverse_BT}
\end{equation}
The compatibility condition of this system turns out to be a copy of (\ref{linearized_universal_hierarchy_eq}) with 
$w$ sub\-sti\-tu\-ted for $v$. Therefore, (\ref{direct_BT}) and (\ref{inverse_BT}) define a B\"acklund 
auto-trans\-for\-ma\-ti\-on for (\ref{linearized_universal_hierarchy_eq}). Since solutions to 
(\ref{linearized_universal_hierarchy_eq}) are identified with local symmetries or shadows of nonlocal 
sym\-met\-ri\-es for Eq. (\ref{universal_hierarchy_eq}), we have

\vskip 10 pt
\noindent
{\sc corollary}. 
{\it
Equations
\[
\left\{
\begin{array}{lcl}
D_x(\psi) &=& u_x  D_t(\varphi)-u_{tx} \varphi + D_y(\varphi),
\\
D_y(\psi) &=& u_y  D_t(\varphi)-u_{ty}  \varphi.
\end{array}
\right.
\]
define a recursion operator $\psi = \mathcal{R}(\varphi)$ for symmetries of the universal hierarchy equation
(\ref{universal_hierarchy_eq}). The inverse recursion operator $\varphi = \mathcal{R}^{-1}(\psi)$  is defined by 
equations
\[
\fl
\left\{
\begin{array}{lcl}
D_t(\varphi)  &=& (D_y(\psi)+u_{ty}\,\varphi)\,u_y^{-1},
\\
D_y(\varphi)  &=&  ((u_y u_{tx}- u_x u_{ty})\,\varphi+u_y \,D_x(\psi) - u_x\, D_y(\psi))\,u_y^{-1}.
\end{array}
\right.
\]
}

\section{Conclusion}
We have showed the possibility to find recursion operators for symmetries of nonlinear {\sc pde}s by means of
Cartan's method of equivalence. While this approach is com\-pu\-ta\-ti\-o\-nal\-ly involved, it does not require any 
preliminary information about linear coverings of the {\sc pde} under study. Its applicability to other {\sc pde}s 
is an interesting problem for the further research.

\section*{Acknowledgments}
I am very grateful to Professors I.S. Krasil${}^{\prime}$shchik, M. Marvan, A.G. Sergyeyev, and M.V. Pavlov for
 valuable  discussions. Also I'd like to thank Professors  M. Marvan and A.G. Sergyeyev for the warm hospitality in 
 Mathematical Institute, Silezian University at Opava, Czech Republic, where this work was initiated and partially 
 supported by the ESF project CZ.1.07/2.3.00/20.0002.

\end{document}